 \documentclass[preprint,aps,prb,showpacs]{revtex4}

\usepackage{epsf}
\usepackage{bm}

\begin{document}

\title {First principle investigation of the structural and electronic properties
of the gallium clusters and their influence on the melting characteristics}

\author{Sailaja Krishnamurty, Kavita Joshi, Shahab Zorriasatein, and D. G. Kanhere}

\affiliation{
Department of Physics, and
Centre for Modeling and Simulation,
University of Pune,
Ganeshkhind,
Pune--411 007,
India}

\begin{abstract}

First principle calculations have been performed to understand the
experimentally observed size sensitive variations in the characteristics 
of heat capacities of gallium clusters [G. A. Breaux {\it et. al.} 
J. Am. Chem. Soc., {\bf 126}, 8628 (2004)]. It was reported that
while some clusters exhibit a clear solid like to liquid like transition others exhibit a
continuous transition with no peak in the heat capacity curve.
In addition, the clusters also exhibit a variation of about 300 K (500--800~K)
in the melting temperature across the size range of 20 to 46. 
In the present work we correlate the observed finite temperature
properties to its geometry and nature of bonding in the ground state.
We demonstrate that the local order (i.e., island of atoms bonded with similar strength)
in the ground state geometry is responsible for the variation in the shape of the
heat capacity curve.
We attribute the higher melting temperature of clusters
to the presence of distinct core and 
strong covalent bonds between the core and surface atoms.

\end{abstract}


\maketitle

\section{Introduction}
\label{sec.intro}

During the past decade or so, a number of systematic calorimetric measurements 
probing the finite temperature behavior of unsupported free clusters have 
yielded unexpected and rich physics.~\cite{pioneer-Na,Melting-clusters-na,Ga-expts,Jarrold-JACS}
One of the first reports is on the measurement of 
melting temperatures (T$_m$) of simple metal clusters of sodium in 
the size range of 55 to 357 atoms.~\cite{pioneer-Na}
The measured melting temperatures were found to be an irregular function of the cluster sizes 
with melting peaks showing no clear correlations with the observed 
magic numbers in sodium.~\cite{Melting-clusters-na} The second measurement was reported by
Jarrold and co--workers which brought out additional important aspects in the melting of finite size clusters.~\cite{Ga-expts,Jarrold-JACS} 
The systems investigated are free clusters of gallium in the size range 
of 30 to 55.~\cite{Jarrold-JACS} The measured heat capacities of the clusters in the above range 
revealed three very interesting features: (1) Higher than bulk melting temperatures (T$_{m[bulk]}$ = 303~K)
in all the clusters (also seen in two smaller clusters viz., Ga$_{17}$ and Ga$_{20}$) in direct 
contradiction with the accepted paradigm, {\it viz.,} the reduction in the melting temperatures 
with the size.~\cite{Ga-expts} (2) The size sensitive behavior of the shape of the heat capacities where 
addition of even one atom results in a dramatic change of shape, prompting some of the clusters to be called as 
``Magic Melters".~\cite{Jarrold-JACS} This means that while some clusters do undergo a conventional and clear melting transition,
 others undergo a near continuous transition making it very 
difficult to identify any meaningful transition temperature. (3) A variation of about 300~K (500--800~K)
in the T$_m$ across the sizes between 20 to 46. 
For example, Ga$_{20}$, Ga$_{46}$, and Ga$_{47}$ exhibit a highest melting temperature of about 800~K while
the intermediate clusters such as Ga$_{31}$, Ga$_{33}$, Ga$_{37}$, Ga$_{41}$, etc., 
exhibit a lower melting temperature of the order of 500--600~K. Thus,
the size sensitive variations in melting temperature are not monotonic in nature. Such variations 
in the melting temperature of clusters are also observed experimentally in the case of
aluminum clusters.~\cite{Al-expt}

An explanation and understanding of some of the unexpected and puzzling experimental observations 
noted above warranted detailed ab initio Molecular Dynamical (MD) simulations based on Density Functional 
Theory (DFT). A clear signal for needing such an approach came from detailed investigations 
using classical inter atomic potentials, which failed to reproduce the data on sodium, 
even qualitatively.~\cite{Na-classical} On the other hand, even early DFT based investigations, 
limited to short simulation times, were successful in providing some insights 
into the phenomena of melting.~\cite{Na-abinitio} Recently, works of 
Chacko et. al.,~\cite{Chacko} and Aguado and Lopez~\cite{aguado} over the entire size range 
have reported excellent agreement with the experimental melting temperatures of these 
clusters. These calculations indicate that the geometry plays a more direct and significant 
role with the electronic structure influencing it in a more subtle way. Indeed it must be 
emphasized that the nature of bonding and the energetics turn out to be two crucial 
ingredients, making an explicit quantum mechanical treatment of electrons mandatory. 

In two short communications, we have addressed the issues of higher than bulk melting point~\cite{Chacko-PRL}
and size sensitive nature observed in gallium clusters.~\cite{Kavita-PRL}
The predominantly covalent nature of bonding in the gallium clusters in contrast with predominant metallic 
bonding in bulk--Ga has been shown to be responsible for the higher than bulk melting point of clusters. 
These reports are limited to Ga$_{13}$ and Ga$_{17}$ which was later on extended to Ga$_{30}$ and Ga$_{31}$. 
By examining the geometry and the finite temperature behavior of Ga$_{30}$ and Ga$_{31}$, the pair 
displaying maximum size sensitivity, we explained that the dramatic change in their heat 
capacities has geometric origin.~\cite{Kavita-PRL} Specifically we established a direct 
correlation between the nature of the ground state and the observed heat capacity. 
An ``ordered" cluster is expected to display a well characterized melting transition 
showing an identifiable, albeit broad peak. On the other hand a completely ``disordered"
cluster will undergo a continuous transition with a very broad heat capacity. 
Quite clearly such a description is qualitative and needs further quantifications. 
It also raises questions about the universality of such a phenomenon. The earlier work 
is based on two specific clusters, viz., Ga$_{30}$ and Ga$_{31}$ (as well as Ga$_{17}$
and Ga$_{20}$)~\cite{Ga1720} where such an effect was discussed and did not address the issue of observed 
systematic behavior over entire range and the shift in T$_m$ to higher value in case of Ga$_{20}$ and Ga$_{46}$.

The present work is an attempt to understand and evolve a complete picture of the experimental
observations over the reported size range. 
Evidently, complete thermodynamic calculations based on an ab initio MD are 
prohibitively expensive. However, our earlier work clearly brings out the two main ingredients 
which are essential and perhaps sufficient for such an understanding, namely the analysis of 
ground state geometry and the nature of bonding in the clusters. In the present 
work, we have carried out such an analysis for the gallium clusters in the size range 
of 20 to 55.
It may be mentioned that many of the geometries do not display any obvious rotational symmetry. 
This is especially true of large clusters. Therefore, it is not straightforward to characterize 
the nature of order and disorder using some quantitative measure. In this sense the 
situation is some what unsatisfactory. However, considerable insight can be gained by 
analyzing the shape of the cluster, distribution of shortest bond lengths (in other words strongest
covalent bonds) and pair correlation function. 
Another extremely useful quantity is the Electron Localization Function (ELF) which provides 
a semi quantitative picture of the nature of bonding and the connectivity of atoms. In view 
of the fact that we have made use of ELF extensively, this function will be discussed in detail later.

It is most profitable to put the present work on gallium clusters in the context of
known structural trends in small gallium clusters and bulk properties. 
The evolutionary trends in the geometries of small clusters ( N $\le$ 26) have been 
investigated by Song and Cao.~\cite{Ga-JCP} In smaller clusters 
it is relatively easy to discern the symmetry. For example, the ground state geometry 
of Ga$_{13}$ is decahedron. The next two structures in size are obtained by capping 
one of the square faces by one and two atoms, respectively. This distorts the 
structure and disturbs the symmetry. By the time four atoms are added, viz., Ga$_{17}$, 
the structure is completely distorted.
This evolutionary trend changes after Ga$_{17}$ and a structure with identifiable 
ordered planes develops e.g., as seen in Ga$_{20}$. Such an evolutionary trend in the geometries 
from well ordered one like Ga$_{13}$ to another partially ordered one like Ga$_{20}$ through a 
disordered one like Ga$_{17}$ is responsible for the size sensitive nature. Since there 
is nothing specific about gallium per se, in so far as the evolutionary trends 
are concerned, the size sensitive nature of the heat capacity should be an universal phenomenon. 
Indeed, our calculation on clusters of Na$_{n}$ (n= 55 and 58),~\cite{Lee-PRL} 
Au$_{n}$ (n=19 and 20),~\cite{Ghazal-Au} and smaller clusters of Ga$_{n}$ (n=17 and 20)~\cite{Ga1720} 
supports this conjecture. In light 
of the above discussion, the availability of experimental heat capacity data on gallium for all the
sizes within a given range presents an interesting opportunity to carry out 
detailed investigations of their ground state properties. 
\begin{figure}
 \epsfxsize=0.50\textwidth
 \centerline{\epsfbox{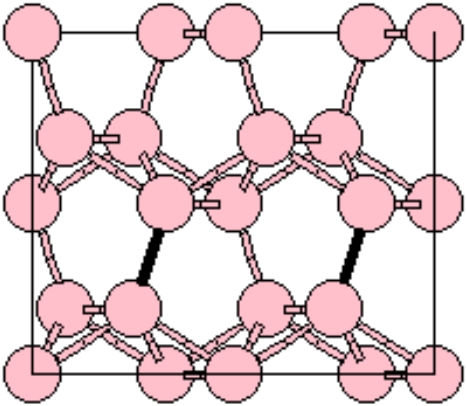}}
 \caption{\label{fig1}
 Unit cell of $\alpha$--Ga.
 It shows two buckled planes. The dark line joining the atoms
 corresponds to the inter planar covalent bond discussed in the text.
 }
\end{figure}

In the bulk phase, gallium has been described as a metallic molecular solid.  
Gong et. al.,~\cite{Gong} have calculated the band structure of $\alpha$-Ga and have 
shown that molecular character and metallic conduction coexist. $\alpha$-Ga is a face 
centered orthorhombic solid with 8 atoms in the cubic unit cell. It has only one nearest 
neighbor at a distance of 2.44~{\AA}. The next three shells contain two atoms each at 
distances of 2.71~{\AA}, 2.74~{\AA}, and 2.83~{\AA}, respectively. The structure can be regarded as 
strongly buckled planes connected by the shortest bonds.
Fig.~\ref{fig1} shows the atomic arrangement from this 
perspective. The short bonds across the planes (dark lines in Fig.~\ref{fig1}) 
show covalent character, while the in plane 
electrons are delocalized. The density of states show a pseudo gap attributed to the covalent 
band. Interestingly, the parent structure called as Ga-II is a distorted tetragonal face centered cubic (FCC),  
which can be obtained under pressure of 35~kbar. This structure is quite close in energy to the 
$\alpha$-Ga and is metallic. In other words Ga-II, the metallic FCC structure rearranges into 
$\alpha$-Ga, once the pressure is released forming a covalent bond. Thus, the tendency to form 
a covalent bond is already there in a nascent form and given a chance to further rearrange the 
atomic positions, as is the case in clusters, the structure may prefer to form covalent bonds. 
This is what we observe in the finite sizes of Ga$_n$ clusters where, n $\approx$ 55. In fact, our detailed analysis
of the ground state geometries of Ga$_n$ (n= 13--55) clusters shows an interesting
pattern as to how the covalent bonds in the cluster rearrange themselves, leading to significant
variations in the melting temperature. At what sizes the partial 
metallic bonds are formed is an interesting question but beyond the scope of this work. 

In what follows, we study the ground state geometries of selected gallium clusters 
(viz., Ga$_n$, n= 13, 17, 20, 30, 31, 33, 37, 46, and 55) and discern the evolutionary trends.
Along with the evolutionary trends, we also analyze the nature of bonding, the number of covalent bonds, 
and their distribution within the clusters. Using these trends, we bring out 
the factors contributing to the non--monotonic variations in the melting temperature. 
Thus, we expect our current work to throw light on general 
experimental observations over the reported size range.

\section{computational details}
\label{sec.comp}

In order to have a realistic guess for the ground state, we have optimized about 300
geometries for each of these clusters. The initial configurations for the optimization were obtained 
by carrying out a constant temperature dynamics of 100~ps each 
at various temperatures between 600 to 1200~K. 
The optimization was carried out using Vanderbilt's ultra soft pseudo potentials~\cite{uspp-vanderbilt}
within the Generalized Gradient Approximation (GGA) for describing the core-valance interactions as
implemented in the \textsc{vasp} package.~\cite{vasp}
For all the calculations, we take the $4s^2$ and $4p^1$ electrons as
valence electrons and the 3$d$ electrons as a part
of the ionic core.
An energy cutoff of about 9.54~Ry is used for the plane wave expansion of 
the wave function, with a convergence in the total energy
of the order of 0.0001~eV.

Once the ground state geometry is obtained from the first principles calculations,
various structural and electronic properties of the cluster in its ground state were analyzed. 
The structural properties analyzed are (i) bond length variations within the cluster,
(ii) distance from Centre of Mass (COM), and (iii) 
shape of the cluster. The shape of the cluster is quantified using the deformation coefficient
($\epsilon_{\rm pro}$). For a given configuration, $\epsilon_{\rm pro}$ is defined as
\begin{equation}
\epsilon_{\rm pro} = \frac{2Q_z}{Q_x+Q_y}
\end{equation}
where, Q$_{x}$, Q$_{y}$ and Q$_{z}$ are the eigenvalues, in ascending order of the quadrapole tensor
\begin{equation}
Q_{ij}=\sum_{I}R_{Ii}R_{Ij}
\end{equation}
{\it I} runs over the number of ions and R$_{Ii}$ is the ith coordinate ({\it i} and {\it j} run from 1
to 3) of the ion {\it `I'} relative to the center of mass of the cluster. 
In simple terms, Q$_{x}$, Q$_{y}$, and Q$_{z}$ define the spread of the cluster
along the X, Y, and Z axis.
Thus, configuration with spherical shape has eigen values Q$_{x}$ $\approx$ Q$_{y}$ $\approx$ Q$_{z}$
($\epsilon_{\rm pro}$=1). A prolate configuration has Q$_z$ $>>$ Q$_y$ $\approx$ Q$_x$,  
while a structure with oblate configuration has Q$_z$ $\approx$ Q$_y$ $>>$ Q$_x$.

The nature of bonding between the atoms in a cluster is analyzed using Electron
Localization Function (ELF).~\cite{elf,Marx} 
For a single determinantal wave function built from Kohn-Sham orbitals, $\psi _{i}$, the ELF is defined as,~\cite{becke}
\begin{equation}
\chi _{{\rm ELF}}=[1+{(D/D}_{h}{)}^{2}]^{-1},
\end{equation}
where
\begin{eqnarray} D_{h}&=&(3/10){(3{\pi
}^{2})}^{5/3}{\rho }^{5/3}, \\ D&=&(1/2)\sum_{i}{\
|{\bm{\nabla} \psi _{i}}|}^{2}-(1/8){|{\bm{\nabla}
\rho }|}^{2}/\rho, \end{eqnarray} with $\rho \equiv \rho
({\bf r})$ being the valence electron density.
D is the excess local kinetic energy density due to
Pauli repulsion and D$_{h}$ is the Thomas--Fermi kinetic energy
density. The numerical values of $\chi _{{\rm ELF}}$ are conveniently
normalized to a value between zero and unity. A value
of 1 represents a perfect localization of the valence charge while the
value for the uniform electron gas is 0.5. Typically, the existence of an
isosurface in the bonding region between two atoms at a high value of $\chi _{{\rm ELF}}$ 
say, $\ge 0.70$, signifies a localized bond in that region.

Recently, Silvi and Savin~\cite{elf} introduced a nomenclature for
the topological connectivity of the ELF. According to this
description, the molecular space is partitioned into regions or
basins of localized electron pairs. At very low
values of ELF all the basins are connected (disynaptic basins). In other
words, there is a single basin containing all the atoms. As the
value of $\chi _{{\rm ELF}}$ is increased, the basins begin to split, and
finally we will have as many basins as the number of atoms.
The value of ELF at which the basins split (a disynaptic basin 
splits into two monosynaptic basins) is a measure
of the interaction between the different basins 
(a measure of the electron delocalization).

\section{Results and discussion}
\label{sec.sim-rd}

In order to gain some insight into the experimentally observed differences in the heat capacity
curves of Ga$_{n}$ ($n=17-55$), we have investigated the ground state
geometries of selected clusters, viz., Ga$_{n}$, $n=13, 17, 20, 30, 31, 33, 37, 40, 44, 46, 55$
and analyzed the differences in their structure and bonding.
The choice of these clusters is dictated so as to represent the changing nature of the heat
capacities across the measured series. For example, among these clusters,
Ga$_{17}$, Ga$_{30}$, and Ga$_{55}$ have very broad heat capacities.
Ga$_{40}$ and Ga$_{44}$ show a weak peak around 550~K and 700~K, 
respectively. 
Ga$_{20}$, Ga$_{31}$, Ga$_{33}$, Ga$_{37}$, and Ga$_{46}$ have a well recognized peak.
Among these, Ga$_{20}$ and Ga$_{46}$ melt around ~800~K, while Ga$_{31}$, Ga$_{33}$, and Ga$_{37}$
melt between 550--600~K. Ga$_{13}$ is taken as a reference cluster to 
analyze the growth pattern in these clusters.   

\begin{figure}
  \epsfxsize=1.00\textwidth
  \centerline{\epsfbox{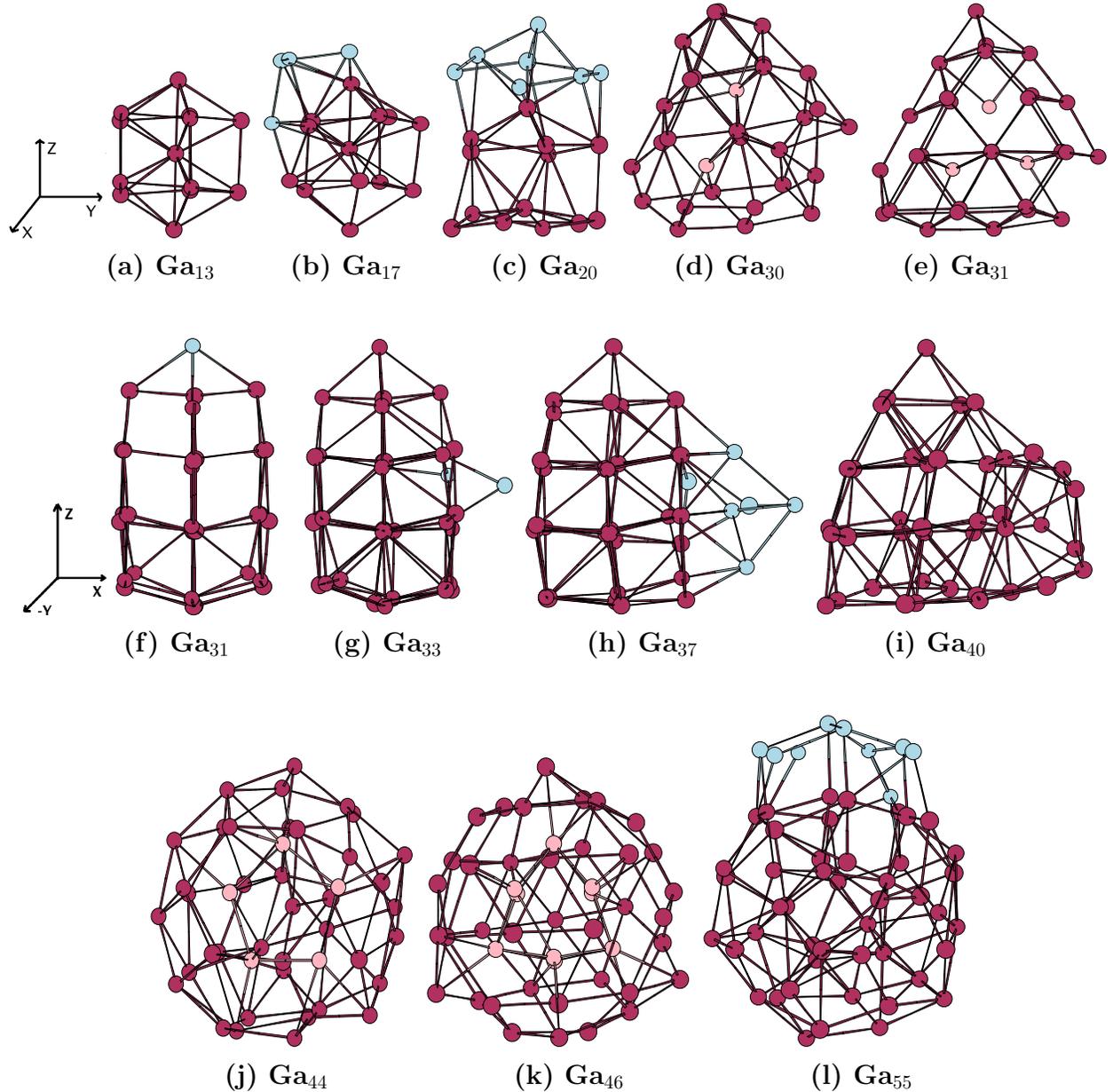}}
  \caption{\label{fig2}
  Ground state geometries of selected gallium clusters between Ga$_{13}$ and Ga$_{55}$  
  }
\end{figure}

\begin{figure}
  \epsfxsize=0.90\textwidth
  \centerline{\epsfbox{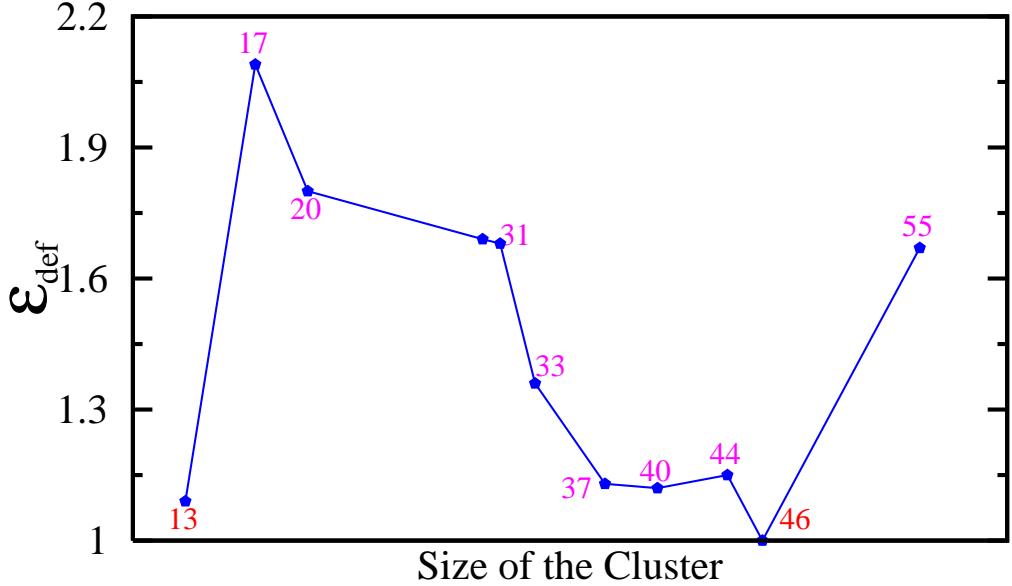}}
  \caption{\label{fig3}
  Deformation co-efficient parameter ($\epsilon$$_{pro}$) of various gallium clusters. 
  }
\end{figure}

\begin{figure}
  \epsfxsize=0.90\textwidth
  \centerline{\epsfbox{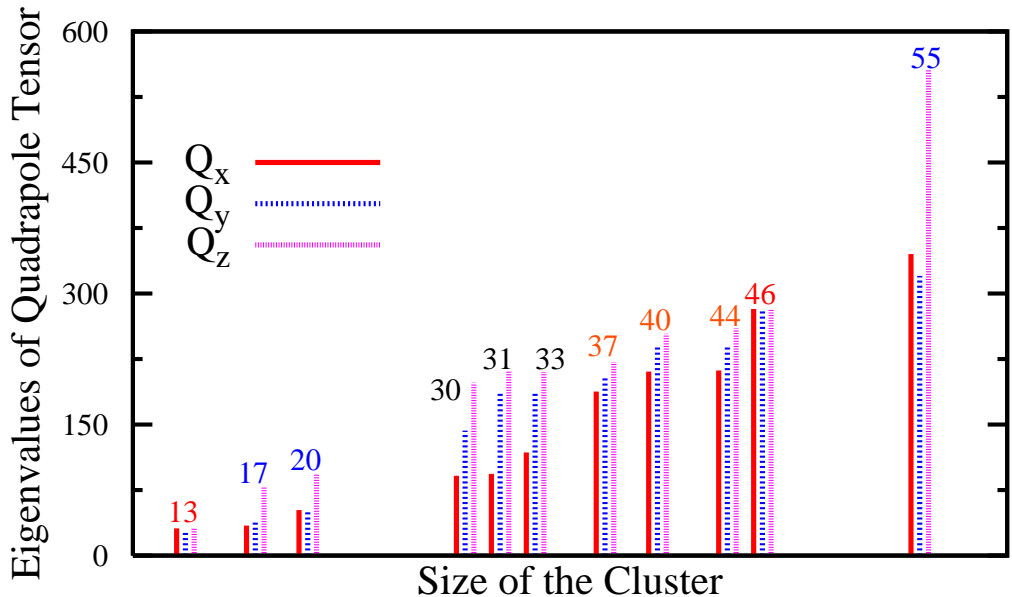}}
  \caption{\label{fig4}
  Eigen values of the quadrapole tensor for gallium clusters.
  }
\end{figure}

In Fig.\ \ref{fig2}, we show the ground state geometry of these clusters obtained in our search.
The atoms shown in light color (blue on line) indicate the growth of the cluster
over the previous one.
The shape of these clusters is analyzed using $\epsilon_{\rm pro}$ 
and eigen values of quadrapole tensor which are shown in Fig.\ \ref{fig3} and Fig.\ \ref{fig4},
respectively.
We begin our discussion with a note on Ga$_{13}$ shown in Fig.\ \ref{fig2}--(a). The ground state geometry of
Ga$_{13}$ was found to be slightly distorted decahedron in previous works.~\cite{ga13,Chacko-PRL,Ga-JCP}
The cluster is nearly spherical in shape with $\epsilon_{pro}$ $\approx$ 1.1.
The inter--planar and core--surface bonds are the shortest ones (2.43--2.65~{\AA}) in the cluster.
The intra--planar bond distances are the longer ones and vary between 2.85--3.00~{\AA}. 
Addition of 4 atoms to Ga$_{13}$ results in a deformation of decahedron as seen from Fig.\ \ref{fig2}--(b).
The atoms shown in red (dark color) indicate
the deformed structure of Ga$_{13}$, while the atoms in blue (light color) show the extra four atoms.
It is clearly seen from this figure as well as from the analysis of Q$_{x}$, Q$_{y}$, and Q$_{z}$ (see Fig.\ \ref{fig4}) that Ga$_{17}$
is a growth over Ga$_{13}$ along Z--axis. The $\epsilon_{\rm pro}$ of Ga$_{17}$ is $\approx$ 2.1. The shortest
bonds (2.50--2.70~\AA) in Ga$_{17}$ are randomly distributed (scattered) within the cluster unlike in
case of Ga$_{13}$, where they are localized. The implication of
this distribution of shortest bonds is brought out more clearly in the later section 
where we discuss their connectivity and its relation with the melting temperature.

Fig.~\ref{fig2}--(c) shows the ground state geometry of Ga$_{20}$. 
The atoms in light color (blue on line) are the extra
seven atoms that form a dome like structure on the deformed
Ga$_{13}$ cluster. The cap to the top plane of Ga$_{13}$
now becomes the core atom in case of the Ga$_{20}$. 
Thus, Ga$_{20}$ is also obtained by addition of atoms on
deformed Ga$_{13}$ along the Z--axis. 
The eigen values of quadrapole tensor and $\epsilon_{\rm pro}$ ($\approx$ 1.8)
clearly indicate Ga$_{20}$ to be prolate in shape.
As in case of Ga$_{13}$ and Ga$_{17}$, the atoms are bonded to each other through bond lengths ranging
between 2.54--3.00~{\AA}. A careful examination of the distribution of bond lengths in  
Ga$_{20}$ cluster reveals that the shortest bond lengths (2.54--2.62~{\AA})
in the cluster are localized in the newly added (top) and bottom planes of the cluster.
The core atom of the cluster is connected to the surface atoms with 
bond lengths of 2.65--2.70~{\AA}, while rest of the atoms within the cluster
are connected to each other through varying lengths of 2.73--3.00~{\AA}.

Addition of 10 atoms to Ga$_{20}$, viz., Ga$_{30}$, results in a configuration with stacked planes as shown in 
Fig.~\ref{fig2}--(d). The Q$_{x}$, Q$_{y}$, and Q$_{z}$ values show that the growth is predominantly
along the Y and Z--axis resulting in a nearly oblate configuration. 
Ga$_{30}$ does not have a distinct core. 
However, it has two interstitial atoms (atoms not facing the surface) connected to each other and to the surface
atoms through bond lengths of $\approx$ 3.0~{\AA} (shown in light red color on line). 
The shortest bonds are randomly distributed within the cluster. Addition of one
atom to Ga$_{30}$ results in a significant reordering of the planes and 
symmetric distribution of atoms in the cluster (cluster has C$_{2v}$ symmetry) 
as seen from Fig.~\ref{fig2}--(e) (for a detailed discussion on Ga$_{30}$ and
Ga$_{31}$ see Ref.~\onlinecite{Kavita-PRL}).
\begin{figure}
  \epsfxsize=0.80\textwidth
  \centerline{\epsfbox{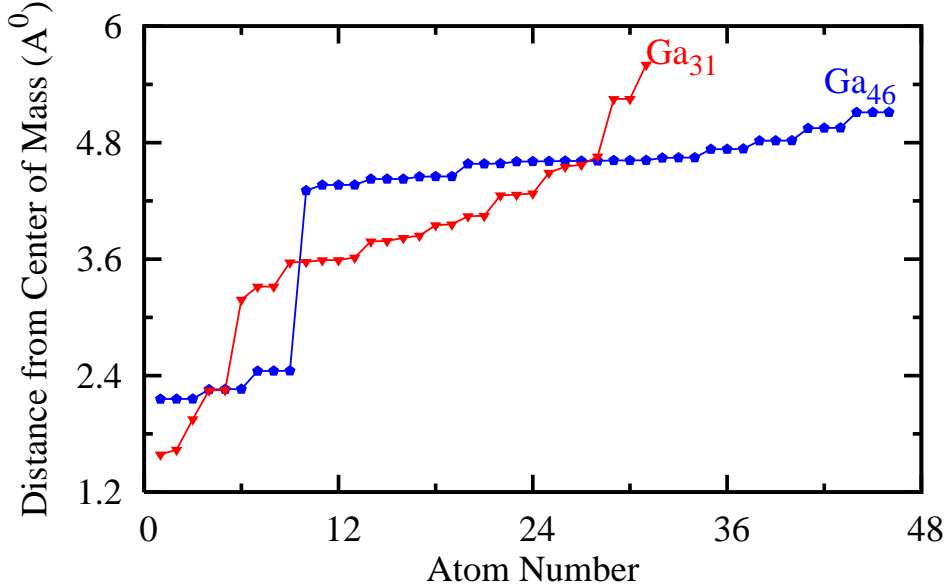}}
  \caption{\label{fig5}
  Distance of atoms from Centre of Mass (COM) in Ga$_{31}$ and Ga$_{46}$ clusters.
  }
\end{figure}
Another perspective of Ga$_{31}$ (Fig.~\ref{fig2}--(f)) and Q$_{x}$, Q$_{y}$, and Q$_{z}$ values clearly show the cluster to be oblate in shape.
Ga$_{31}$ also does not have a distinct core. 
It has three interstitial atoms (shown in light red color on line) which are weakly bonded (distances $\ge$ 2.90~\AA) 
to each other as well as to surface atoms. The shortest bonds in the cluster are localized within the same
plane. 
Fig.~\ref{fig2}--(g) shows Ga$_{33}$ along the X--Z plane. We note that while
Ga$_{33}$ remains identical to Ga$_{31}$ along the Y--Z plane (figure not shown),
what is visibly different is the presence of an extra protruding atom 
along the X--axis. This growth continues in Ga$_{37}$
as seen from its ground state geometry 
in Fig.~\ref{fig2}--(h). This growth is also evident from the increasing values of Q$_{x}$ and a
sudden decrease in the $\epsilon_{\rm pro}$. 
Both Ga$_{33}$ and Ga$_{37}$ retain the stacked planes configuration similar to Ga$_{30}$ and Ga$_{31}$.
The planes are well defined in both the cases and the clusters do not have a distinct core.
The clusters continue to grow along the the X--axis as seen in the
ground state geometry of Ga$_{40}$ (shown along the X--Z plane) in Fig.~\ref{fig2}--(i).
Addition of four more atoms to Ga$_{40}$, however, leads to a significant rearrangement of atoms within the cluster
with out any preferential growth along any axis resulting in a nearly
spherical structure of Ga$_{44}$ (shown in Fig.~\ref{fig2}--(j)) with $\epsilon_{\mathrm{p}}$ $\approx$ 1.1.
We note that as the cluster evolves into a spherical shape as in case of Ga$_{44}$, we once again note the
presence of a distinct core shell. Ga$_{44}$ has five core atoms (shown in light red color on line). 
Addition of 2 atoms to Ga$_{44}$ results in a highly symmetrical cluster (C$_{3v}$ symmetry) with 9 core atoms 
and a spherical configuration (Ga$_{46}$ shown in Fig.~\ref{fig2}--(k)) with
$\epsilon_{\mathrm{p}}$ $\approx$ 1.0. To summarize, the clusters
undergo a spherical to prolate transition (Ga$_{13}$--Ga$_{20}$) followed 
by a prolate to oblate transition (Ga$_{20}$--Ga$_{31}$) and an oblate
to spherical transition (Ga$_{31}$--Ga$_{46}$). As the clusters grow further, 
it is seen that the atoms begin to accumulate
once again preferentially along one axis, leading to prolate configurations as seen in Ga$_{55}$ (Fig.~\ref{fig2}--(l)).
We predict here that
the clusters are likely to undergo once again a prolate to oblate and an oblate to spherical transition as the number of
atoms continue to increase. In the studied size range we notice that the clusters with spherical configurations have
distinct core, while the clusters with oblate configurations do not have any core. This is clearly seen 
from Fig.~\ref{fig5}, in which we show the
distance of atoms from the COM of Ga$_{31}$ which is oblate in shape and Ga$_{46}$, a 
spherical structure. Ga$_{46}$ has a clear shell structure, while Ga$_{31}$
does not have a distinct core but is characterized by interstitial atoms.

\begin{figure}
  \epsfxsize=1.00\textwidth
  \centerline{\epsfbox{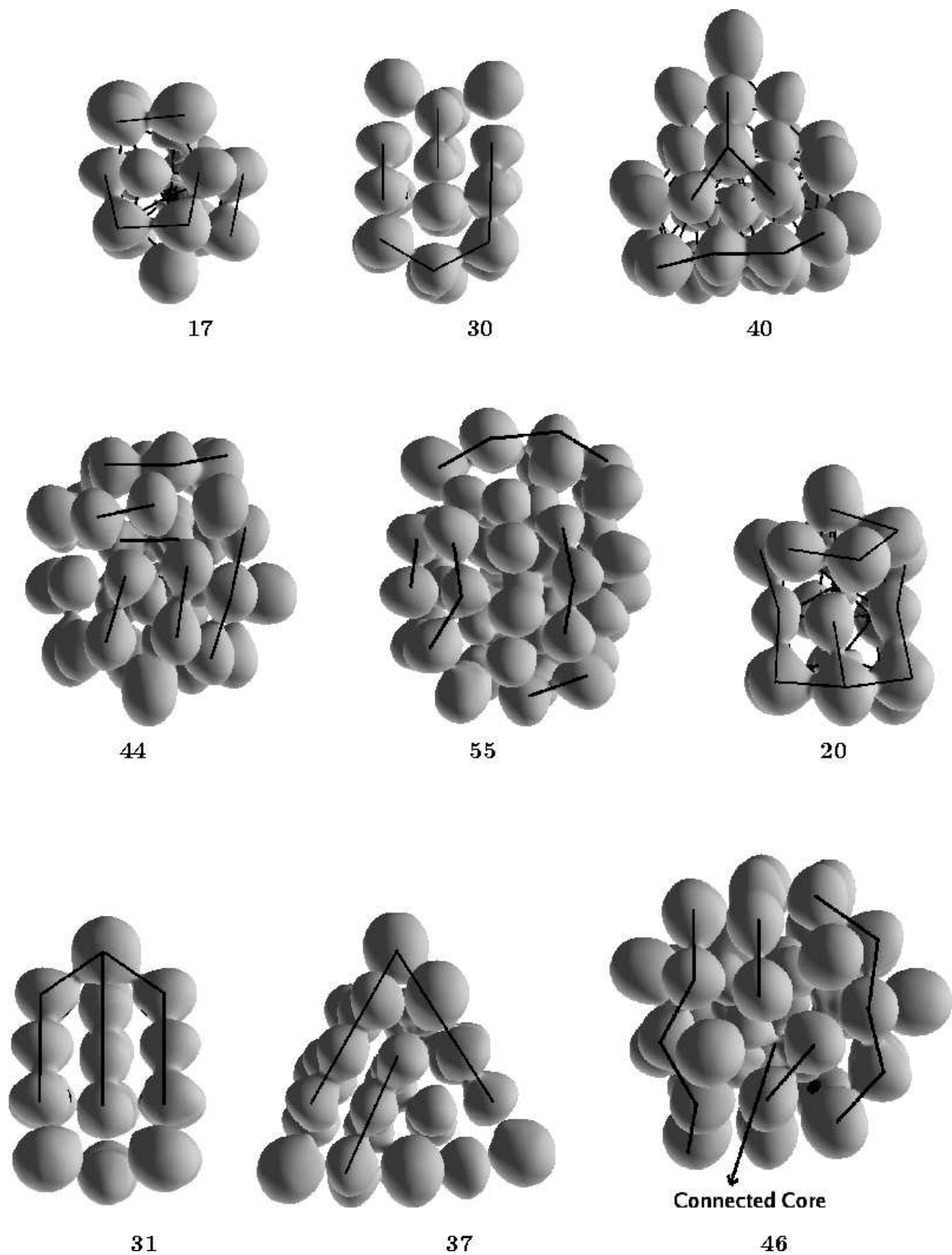}}
  \caption{\label{fig6}
  Electron Localization Function (ELF) at an isovalue of 0.70 for various gallium clusters.
  }
\end{figure}

To understand the reasons behind the occurrence of so called ``melters" and ``non--melters", 
we analyze the nature of bonding in the clusters studied. 
In Fig.~\ref{fig6}, we show the isosurface of ELF at a value of $\chi _{{\rm ELF}}$ $=$ 0.70 for all the clusters.
The black lines join the atoms in the same basin.
A careful examination of the basin structures reveals an interesting contrast between 
``non--melters" (Ga$_{n}$, n= 17, 30, 40, 44, and 55) and ``melters" (Ga$_{n}$, n=20, 31, and 37).
The ``non--melters" show a fragmented pattern of basins. In other words, there are
many basins each consisting of not more that 4--5 atoms. In contrast, 
``melters" have at least one large basin consisting of 10 atoms or more. 
Ga$_{46}$, a ``melter", apparently appears to have fragmented basin as seen visually. 
A detailed examination reveals that all the 9 core atoms 
are connected at the value of $\chi _{{\rm ELF}}$ = 0.70 forming a single basin (indicated by an arrow). We further note that these core atoms
are also connected to the surface atoms, there by forming a single large basin of 15 atoms. This
is made more clear in the next section where we show the schematic representation of the covalent bonds
in each cluster. 

To summarize, the clusters exhibiting a sharp solid--liquid transition 
are seen to have substantial number of atoms connected through a single basin. 
Thus, a large group of atoms are bonded together with a similar
strength forming an island of local order and therefore it is reasonable to expect that they will
`melt' together. In this sense the cluster can be considered as (at least partially) ordered
and will show a well defined peak in the heat capacity. 
On the contrary, in clusters exhibiting broad or continuous solid--liquid transition, each atom (possibly a group of atoms)
has different local environment.
That means different atoms are bonded with the rest of the system with varying strength.
Consequently, their dynamical behavior as a response to temperature will differ.
Some of the atoms pickup kinetic energy at low temperatures, while the others may do so at higher
temperatures. Therefore, we can expect the cluster to have
a broad and continuous melting transition. 
Thus, our results confirm our earlier argument
that there exists a definitive relationship between the local order
in the cluster and its finite temperature behavior (and consequently the characteristics of the heat--capacity curve).~\cite{Kavita-PRL} 

We also show that as the cluster grows in size,
it evolves through a succession of ordered and disordered geometries. In such
cases an addition of one or few atoms changes the nature of
the ground state geometry abruptly. The variation in the ground state results in the appearance or
disappearance of the local order within the structure, leading to presence or absence of the melting peak,
respectively. We now realize that this size sensitive nature of heat capacities
is generic to small clusters and related to the evolutionary pattern 
seen in their ground states. The evidence for this
comes not only from gallium clusters but also from clusters 
of sodium,~\cite{Lee-JCP,Lee-PRL} aluminum,~\cite{Kavita-PRB} and gold~\cite{Ghazal-Au} having
very different nature of bonding.
\begin{figure}
  \epsfxsize=0.90\textwidth
  \centerline{\epsfbox{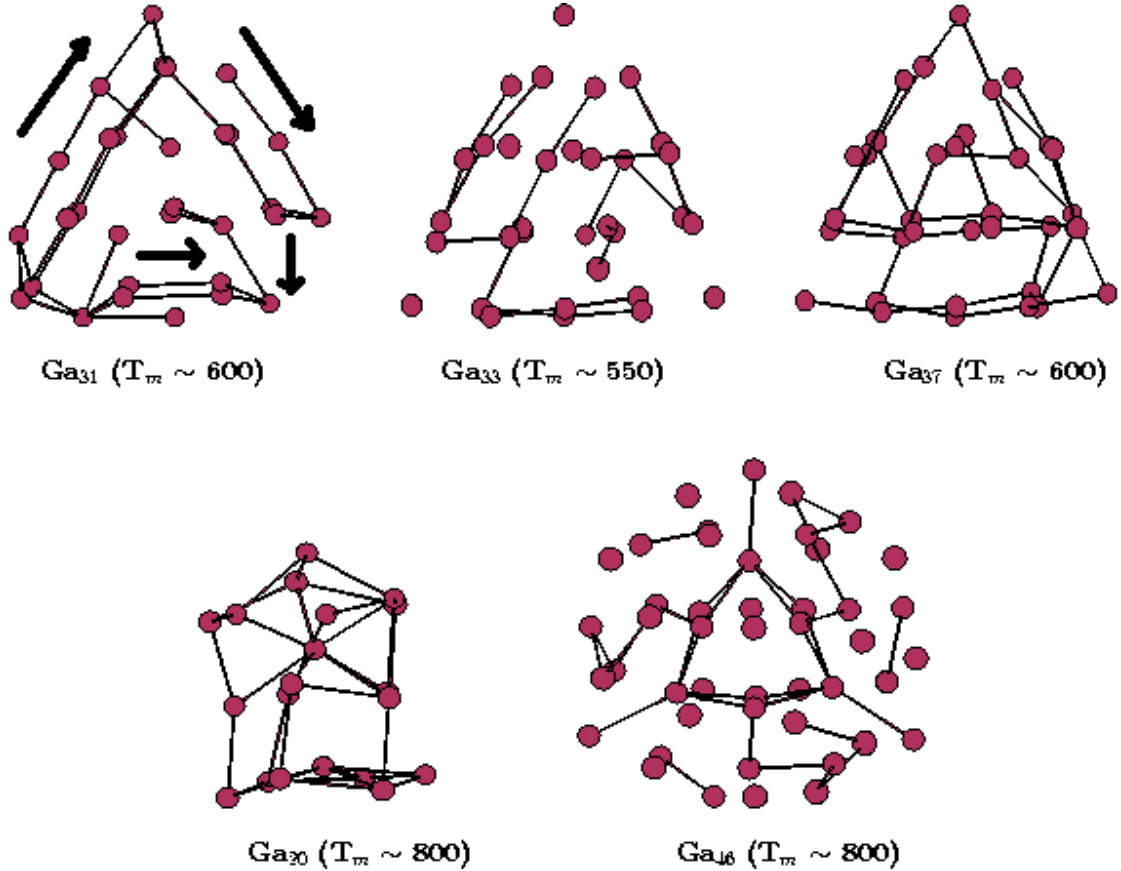}}
  \caption{\label{fig7}
  Distribution of shortest bonds within the gallium clusters. 
  }
\end{figure}
\begin{figure}
  \epsfxsize=0.80\textwidth
  \centerline{\epsfbox{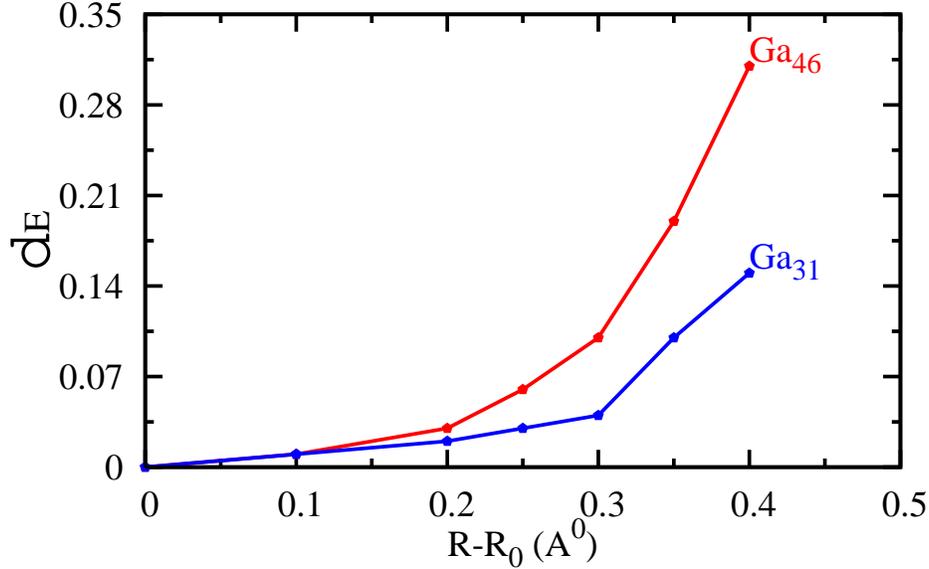}}
  \caption{\label{fig8}
  Variation in the binding energy ({\bf d}$_{E}$) of cluster with respect to the displacement of surface atoms from their equilibrium positions (R$_{0}$). 
  }
\end{figure}

Next, we attempt to understand the reason behind the variations in the melting temperature
in Ga$_{n}$ (n = 20--46) clusters.
We analyze the structure and bonding of the clusters exhibiting a distinct melting transition viz., Ga$_{13}$,
Ga$_{20}$, Ga$_{31}$, Ga$_{33}$, Ga$_{37}$, and Ga$_{46}$. It may be recalled that Ga$_{20}$ and Ga$_{46}$ melt
around 800~K, while Ga$_{31}$, Ga$_{33}$, and Ga$_{37}$ melt between 500--600~K. The first striking difference
to note is the presence of core atoms in the Ga$_{20}$ and Ga$_{46}$ and absence of any distinct core in
Ga$_{31}$, Ga$_{33}$, and Ga$_{37}$. The distribution of
the shortest bonds ($\le$ 2.70~{\AA}) in each cluster 
(which are also the atoms forming a single basin as indicated by ELF) is shown in Fig.\ \ref{fig7}. 
It is clearly noted that the shortest bonds 
in Ga$_{20}$ and Ga$_{46}$ are spread through out
the cluster and more importantly they make a network connecting the
core and the surface atoms. This results in a compact and
stable electronic configuration in both the clusters as indicated by
bunched eigen value spectrum as compared to evenly spread 
spectrum in case of all other clusters (figure not shown). 
On the other hand, in case of Ga$_{31}$, Ga$_{33}$, and Ga$_{37}$
the covalent bonds are restricted along the planes of the cluster. 
Thus, the pattern of bonding within these three clusters appears 
similar to that of pillared materials, where the atoms along the same plane
are covalently bonded, while the bonds across the planes are 
weakly covalent or more metallic like in nature.
Hence, when these clusters are heated, the weaker inter--planar bonds
are the first ones to break. This is followed by sliding of planes
(constituting of equivalently bonded atoms) along each other.  
This argument correlates well with our analysis of ionic motion in
Ga$_{31}$,~\cite{Kavita-PRL} where we see
that as the cluster is heated, the planes begin to slide
as shown by arrows in Fig.~\ref{fig7}.
This is initiated at a fairly lower temperature of 400~K.
On further increase in the temperature of the cluster, 
the sliding planes collide against each other and
the cluster melts around 600~K.
On the other hand, the strong bonds between the core and surface atoms in Ga$_{20}$ hold the surface atoms
around their respective positions until a temperature of 600~K.~\cite{Ga1720} The cluster melts finally
around 800~K. We further note that another cluster with
a distinct core viz., Ga$_{13}$ also exhibits a network of strong covalent bonds between
the core and surface atoms. Our earlier simulations indicate this cluster to melt at a relatively 
high temperature of 1300~K.~\cite{Chacko-PRL}

Thus, we believe
that the presence of a network of strong covalent bonds
between the core and surface atoms results in greater stabilization of
the surface atoms in Ga$_{20}$ and Ga$_{46}$. These surface atoms 
therefore need to overcome a higher energy barrier to diffuse across
the surface as compared to one in Ga$_{31}$, Ga$_{33}$, and Ga$_{37}$. 
To demonstrate this, we stretch one of the atoms
on the surface of Ga$_{31}$ and Ga$_{46}$.
This is done so as to mimic the motion of the atoms when the
cluster is heated. The change in the binding energy of the cluster as function
of the bond stretch is plotted in Fig.~\ref{fig7}. It is easily noted
from the figure that the surface atom of Ga$_{46}$ requires more energy for displacement
from its equilibrium position as compared to the one in Ga$_{31}$. 
This is true for most of the surface atoms of Ga$_{46}$ and Ga$_{31}$. 
This analysis is also found to be similar in case of aluminum and sodium clusters. 
Our understanding is also in agreement with the results of Aguado and Lopez~\cite{aguado} for sodium clusters,
where they have found a strong correlation between variation in T$_m$ and core--surface distances.

\section{Summary and conclusion\label{sec:concl}}

In the present work we have attempted to explain the reasons behind the
characteristic features observed in the experimental heat capacity
curves of gallium clusters in the size range of 20--55.
As the gallium clusters grow, they are seen to pass through a cycle of
spherical--prolate, prolate--oblate, and oblate--spherical transition.
During the process they pass through a succession of geometries with
and without local ``order" in bonding. It is the presence of a local ``order"
in the cluster that is responsible for the sharp peak in 
the heat capacity curve. Our studies also show that it is the presence of
a distinct core and more importantly presence of a network of
strong bonds between the core and surface atoms that is responsible
for the higher melting temperature of the clusters.
\section{acknowledgments}

We acknowledge the discussions and help received from Sharan Shetty during
the search for ground state isomers. 
We acknowledge C-DAC(PUNE) for providing us
with supercomputing facilities. KJ and DGK thank Indo French Center For
Promotion of Advanced Research (IFCPAR) for partial financial support (Project No. 3104-2).

\end{document}